\documentclass[sigconf,anonymous=false]{acmart}

\usepackage{tikz} 
\usetikzlibrary{arrows.meta, positioning}
\usepackage{booktabs}
\usepackage{tabularx}
\usepackage{array}
\usepackage{ragged2e}

\AtBeginDocument{%
  \providecommand\BibTeX{{%
    \normalfont B\kern-0.5em{\scshape i\kern-0.25em b}\kern-0.8em\TeX}}}

\setcopyright{acmcopyright}
\copyrightyear{2018}
\acmYear{2018}
\acmDOI{10.1145/1122445.1122456}

\usepackage{booktabs} 
\usepackage{url}

\usepackage{caption}
\usepackage[all]{nowidow}
\usepackage{wrapfig}
\usepackage{array}
\usepackage{arydshln}
\usepackage{tabularx}
\usepackage{multirow}
\usepackage{arydshln}
\newcolumntype{L}[1]{>{\raggedright\let\newline\\\arraybackslash\hspace{0pt}}m{#1}}
\newcolumntype{C}[1]{>{\centering\let\newline\\\arraybackslash\hspace{0pt}}m{#1}}
\newcolumntype{R}[1]{>{\raggedleft\let\newline\\\arraybackslash\hspace{0pt}}m{#1}}

\usepackage{wrapfig,lipsum,booktabs} 

\def\authnotes{1}
\newcounter{notectr}[section]
\newcommand{\thenote}{\thesubsection.\arabic{notectr}\refstepcounter{notectr}}

\newcommand{\note}[2]{$\ll$#1~\thenote: #2$\gg$}
\newcommand{\cnote}[1]{\ifnum\authnotes=1 \textcolor{blue}{\note{Comment:}{#1}}\fi}

\copyrightyear{2026}
\acmYear{2026}
\setcopyright{acmlicensed}\acmConference[ACM E-Energy'26]{E-Energy}{Nov 5-9, 2026}{USA}
\acmBooktitle{E-Energy'25, Nov, 2026, USA}
\acmPrice{15.00}
\acmDOI{10.1145/3491asdhgkfdj}
\acmISBN{978-1-4503-9157-3/22/04}

\begin{document}



\title[Data Center Audit]{Beyond PUE: A Local-Impact Audit Framework for Data Center Environmental Accountability}


\author{Sharifa Sultana}
\affiliation{
  \institution{University of Illinois Urbana-Champaign}
  \country{USA}}
\email{sharifas@illinois.edu}

\author{Syed Ishtiaque Ahmed}
\affiliation{%
  \institution{University of Toronto}
  \country{Canada}}
\email{ishtiaque@cs.toronto.edu}

\renewcommand{\shortauthors}{Sultana \& Ahmed}

\begin{abstract}
Standard data center sustainability metrics, including Power Usage Effectiveness (PUE), Water Usage Effectiveness (WUE), and Carbon Usage Effectiveness (CUE), measure a facility's resource use and emissions intensity, normalized to IT energy use, without directly representing local resource scarcity, infrastructure capacity, or social footprint. This gap has become politically consequential. In the first quarter of 2026 alone, local opposition delayed or canceled roughly \$130 billion in projects across the United States, driven overwhelmingly by concerns over amid recurring concerns over water use, power demand, infrastructure capacity, and transparency rather than internal efficiency, matching the total for all of 2025 \cite{datacenterwatch2026q1}. We propose a five-category local-impact audit framework covering efficiency, water stewardship, carbon and renewables, regulatory compliance, and local disclosure. The framework is designed for recurring quarterly assessment and independent verification against public records. We illustrate its application using publicly available data from three Illinois facilities that are currently at the center of local policy disputes, and we examine the data-access barriers that constrain independent verification. We position this framework as both a research contribution and a practical instrument for county-level policymakers evaluating data center permitting and moratorium decisions.

\end{abstract}



\settopmatter{printfolios=true}

\maketitle

\vspace{-10pt}
\section{Introduction}

Since 2024, the buildout of AI-driven data center infrastructure has collided with local politics in an unusually sharp way. Nationally, local opposition contributed to the delay or cancellation of an estimated \$156 billion in data center projects in 2025 alone~\cite{datacenterwatch-q3q4-2025}, and by mid-2026, organized resistance had spread to at least 833 activist groups across 49 states~\cite{datacenterwatch2026q1}. Illinois offers a concentrated illustration of this trend. In April 2026, the Champaign County board enacted a year-long moratorium on new data center development in direct response to resident concerns over projected energy and water demand~\cite{routefifty2026}. Neighboring municipalities faced related water-infrastructure pressures: DeKalb approved new facilities as residents faced rising water rates, while Joliet approved development during an already-underway transition from a depleted aquifer to Lake Michigan water~\cite{shawlocal2026}. At the state level, the proposed POWER Act would have mandated quarterly water-use reporting and renewable energy sourcing for hyperscale facilities, but failed to pass before the 2026 legislative session ended~\cite{thorsberg2026illinois}.

A recurring theme across these disputes is that the metrics data center operators typically report, Power Usage Effectiveness (PUE), Water Usage Effectiveness (WUE), and Carbon Usage Effectiveness (CUE), answer a narrower question than the one communities and regulators are actually asking. These metrics describe how efficiently a facility converts input resources into IT energy use; they say almost nothing about what the facility takes from the specific place it occupies, or what local capacity exists to absorb that demand. A facility with an excellent PUE can simultaneously draw potable water from an already-stressed aquifer, or add load to a power grid with limited local headroom, while reporting numbers that look identical to a facility with negligible local impact. Indeed, an estimated 40\% of U.S. data centers are sited in regions of high or extreme water stress~\cite{hegde_data}, meaning this blind spot is not a hypothetical concern but a structural feature of where the industry is currently building.

This paper makes two contributions. First, we propose a five-category audit framework, spanning energy efficiency, water stewardship, carbon and renewable sourcing, regulatory compliance, and local disclosure, designed explicitly to surface local, place-based impact rather than facilities' resource-use and emissions intensity alone, and structured for recurring (e.g., quarterly) application rather than one-time certification. Second, we apply this framework to publicly available data from three Illinois facilities currently at the center of active policy disputes, both to illustrate the methodology and to document, empirically, how much of the framework's input data is \emph{not} currently available without a formal, contracted audit relationship. We argue this second finding is itself significant: the same transparency gap that frustrates residents and regulators also limits what any independent researcher can verify from public records alone, and motivates the case for third-party audit mechanisms of the kind we propose here.
\vspace{-10pt} \section{Related Work and the Measurement Gap}

Power Usage Effectiveness (PUE), introduced by The Green Grid~\cite{greengrid-wp49}, is the industry's dominant metric; the Uptime Institute's 2024 survey reports an average of 1.56, flat for five years, with only 41\% of operators tracking water use at all~\cite{uptime2024survey}. PUE's methodology has drawn academic scrutiny~\cite{ieee-pue-measurement, axpue2013, uptime-pue-analysis2023}. WUE and CUE extend the same ratio structure to water and carbon~\cite{patterson2011wue, belady2010cue}; ERE credits heat reuse~\cite{patterson2010ere}; harmonization efforts span these metrics internationally~\cite{harmonizing2011}.

Certification frameworks go further: LEED for Data Centers~\cite{usgbc-leed-dc} and ENERGY STAR~\cite{epa-energystar-dc} (only 270 of 5,381 U.S. facilities certified~\cite{hwglaw2024}) evaluate building-level criteria; the EU Code of Conduct promotes best practices~\cite{jrc2024eucoc}; CDP aggregates self-reported disclosures~\cite{cdp-overview}. Some require periodic renewal, ENERGY STAR recertifies annually via a licensed engineer's site visit, and EU Code of Conduct participants report energy data annually, but none independently verify actual operational consumption at the cadence this paper proposes; LEED's building-level award, once granted, is not automatically revisited as practices change.

A smaller literature engages local impact directly. Mytton documents 1.7 billion liters/day of U.S. data center water use, over half from potable sources, with fewer than a third of operators measuring it~\cite{mytton2021water}; Macknick et al.\ provide foundational water-intensity data for electricity generation~\cite{macknick2011}. Most relevant here, Li et al.\ weight AI workload placement by local water stress~\cite{li2023thirsty, li2024equitable}, and Wu et al.\ propose a water-stress-weighted computing metric~\cite{li2025waterstress}, both applying location-sensitivity to workload placement rather than facility-level audit, a distinction we return to in Section 3. We draw on WRI's Aqueduct atlas for the underlying water-stress data~\cite{wri-aqueduct}; a recent survey documents the broader shift toward context-aware AI infrastructure metrics~\cite{unified-metric-2025}.

What existing instruments share is the evaluation of efficiency independent of location: a PUE-1.1 facility drawing from a depleted aquifer looks identical to one recycling wastewater elsewhere. This is compounded by a reliance on operator self-reporting, evident even in the workload-placement literature just discussed, which improves on location-blindness but still assumes the underlying consumption figures are accurate and complete. Existing frameworks partially address these individually; ENERGY STAR and the EU Code of Conduct both recur annually, for instance, but none combine recurring reapplication with location-sensitivity, independently verified inputs, and a clear separation of compliance from genuine stewardship in a single instrument. Section 3 addresses that combination directly, through context-adjusted weighting, compliance gating, a verification maturity ladder, and trend tracking.

\vspace{-12pt}\section{Proposed Audit Framework}

A framework built only from Section 2's critique, add the categories existing metrics omit, risks producing a checklist rather than an instrument capable of discriminating between facilities in a way that survives scrutiny from operators, regulators, and communities simultaneously. We therefore propose not just a set of weighted categories, but four design principles intended to make the framework more resistant to the specific failure modes we observed while reviewing the Illinois cases discussed in Section 4: static weights that ignore local context, linear aggregation that lets strong performance in one category mask serious failure in another, an undifferentiated treatment of data quality, and a single-snapshot score that cannot represent a facility's trajectory over time.

\textbf{Base categories.} At its foundation, the framework scores five categories, each 0--100 (see Table.1): energy efficiency, water stewardship, carbon and renewable sourcing, regulatory compliance, and local disclosure. Water stewardship and local disclosure together carry 45\% of the base weight, exceeding energy efficiency and carbon/renewables combined (40\%), because these are precisely the categories our review of Illinois disputes found driving community and regulatory concern, while remaining unaddressed by PUE, WUE, and CUE in isolation. Regulatory compliance is weighted lowest (15\%) deliberately: holding a clean compliance record is a floor, not a signal of stewardship, since a facility can withdraw the maximum its permit allows and still register zero violations.

\begin{table}[t!]
\centering
\caption{Proposed audit framework categories and weights} \vspace{-10pt}
\renewcommand{\arraystretch}{1.0}
\begin{tabular}{
  >{\raggedright\arraybackslash}p{1.5cm}
  >{\raggedright\arraybackslash}p{0.4cm}
  >{\raggedright\arraybackslash}p{5.55cm}
}
\toprule
\textbf{Category} & \textbf{Wt.} & \textbf{Metrics / Data Sources} \\
\midrule
Energy efficiency & 20\% & PUE, ERE (heat reuse); operator-reported or metered \\
\addlinespace
Water stewardship & 25\% & WUE, withdrawal vs. permit cap, source type (potable/reclaimed/seawater), \% returned as wastewater; state water permits, utility billing records \\
\addlinespace
Carbon / renewables & 20\% & CUE, renewable \% (annual REC-matched vs. 24/7 hourly-matched, weighted higher); utility interconnection filings, RE contracts \\
\addlinespace
Regulatory compliance & 15\% & Permit violations, NPDES discharge filings, fines; state environmental agency records \\
\addlinespace
Local disclosure \& impact & 20\% & Groundwater level trend near site, logged community complaints, timeliness/accuracy of public reporting; USGS groundwater monitoring, county records, local news \\
\bottomrule
\end{tabular}\vspace{-15pt}
\end{table}

\textbf{Principle 1: Context-adjusted weighting.} A fixed weighting scheme, however well justified in the abstract, treats a facility in a water-abundant region identically to one drawing from an aquifer already projected to fail peak demand, so long as both report the same raw WUE. We propose instead that the water stewardship weight be adjusted upward as a function of a recognized regional water-stress index (e.g., the World Resources Institute's Aqueduct tool)~\cite{wri-aqueduct}, and that the carbon/renewables weight be adjusted as a function of local grid carbon intensity and available headroom. Under this scheme, identical operational numbers can yield different composite scores depending on where the facility sits, extending to facility-level audit the location-sensitivity that recent computer-systems work has begun to apply to workload placement~\cite{li2024equitable, li2025waterstress}, a property missing from every certification framework surveyed in Section 2, and one our Yorkville and Sangamon County case studies suggest local stakeholders already reason about informally, even though no existing audit instrument encodes it formally.

\textbf{Principle 2: Compliance gating, not just weighting.} A linear weighted sum allows a facility with an outstanding water stewardship score to compensate for a serious, active permit violation, an outcome we consider indefensible for an instrument meant to inform permitting decisions. This concern is not merely theoretical: analysis of six major commercial ESG rating agencies found their scores diverge substantially, with much of that divergence attributable to inconsistent measurement and unweighted aggregation of unevenly verified inputs rather than genuine disagreement about underlying performance~\cite{berg2022aggregate}. We therefore propose a gate: any unresolved major regulatory violation caps the composite score at a fixed ceiling (we suggest 40, subject to further calibration) regardless of performance elsewhere, converting compliance from one input among five into a threshold condition the facility must clear before the other four categories are meaningfully considered. 

\begin{figure}[!t]
\centering
\vspace{-10pt}    \includegraphics[width=0.6\columnwidth]{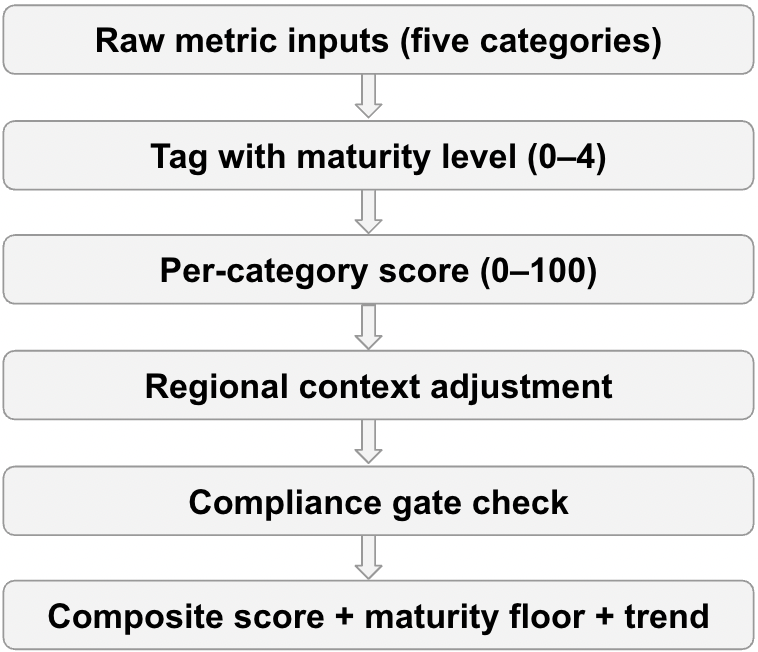} \vspace{-10pt}
    \caption{Proposed scoring pipeline: from raw inputs to a composite score reported with an explicit maturity floor and trend indicator.} \vspace{-20pt}
    \label{fig:figure1}
\end{figure}

\textbf{Principle 3: A verification maturity ladder.} Our case studies exposed a spectrum of data quality that a three-way self-reported/ verified/estimated flag only partially captures. We propose a five-level maturity ladder for each metric, adapting the graded-level structure long used in software process assessment~\cite{paulk1993cmm} to environmental data verification (see Table-2): Level 0, no data publicly available; Level 1, operator-stated projection with no enforcement mechanism (Yorkville's early Project Cardinal figures); Level 2, a permitted cap with contractual penalties for exceedance, but no independent metering (Project Cardinal's current status); Level 3, third-party metering by a public utility, verifying compliance with a cap but not disclosing site-specific totals (DeKalb's arrangement with Meta, whose sustainability reporting publishes only company-wide aggregate water figures, not per-facility consumption); and Level 4, continuous or scheduled telemetry access granted directly to an independent auditor under contract, the level this paper's proposed quarterly-audit mechanism is designed to reach. A composite score should be reported alongside its lowest-maturity contributing input, since a facility scoring well on four Level-4 metrics and poorly on one Level-0 estimate should not be presented with the same confidence as one scoring identically across five Level-4 metrics.

\textbf{Principle 4: Trajectory over snapshot.} Because this framework is designed for quarterly reapplication rather than one-time certification, each quarter's composite score should be reported alongside a trend indicator (improving, stable, or degrading) relative to the trailing four quarters. A facility with a mediocre but steadily improving water stewardship score presents a different risk profile, to an investor and to a permitting authority alike, than one with an identical current score that has been degrading for a year; the composite alone erases that distinction, while a paired trend indicator preserves it at negligible added complexity.

Fig.~1 summarizes the pipeline: inputs are maturity-tagged, scored, context-adjusted, gated, and reported as a composite paired with a maturity floor and trend indicator, never a single undifferentiated number. This requires real implementation effort, contextual data, gate calibration, and audit access for higher maturity levels, but each principle closes one way a composite score can be misleading: masking a compliance failure, hiding an unverifiable input, or flattening a degrading trend into a snapshot of adequacy. We treat that cost as the price of an instrument intended for real permitting and investment decisions, not a defect to be minimized away.

\vspace{-5pt}\section{Illustrative Application: Three Illinois Facilities}
\begin{table}[t!]
\centering
\vspace{-10pt}\caption{Proposed audit framework categories and weights} \vspace{-10pt}
\renewcommand{\arraystretch}{1.0}
\begin{tabular}{
  >{\raggedright\arraybackslash}p{0.3cm}
  >{\raggedright\arraybackslash}p{7.2cm}
}
    \toprule
    \textbf{Lvl} & \textbf{Description} \\
    \midrule
    0 & No data publicly available \\
    1 & Operator-stated projection, no enforcement mechanism \\
    2 & Permitted cap with contractual penalties, no independent metering \\
    3 & Third-party metering (e.g., municipal utility), cap verified but totals undisclosed \\
    4 & Continuous/scheduled telemetry access under independent audit contract \\
    \bottomrule
  \end{tabular}\vspace{-15pt}
\end{table}


To illustrate the framework and probe its practical limits, we apply it to three Illinois facilities selected for the volume of public documentation surrounding their approval processes: Project Cardinal in Yorkville, the Meta data center in DeKalb, and the CyrusOne facility in Sangamon County.

\textbf{Project Cardinal (Yorkville).} This 1,000-acre, 14-building campus illustrates the framework's water stewardship category unusually well, precisely because its reported water figures have shifted substantially over time. Developers initially projected consumption of 500,000 gallons per day; by a September 2025 filing this had fallen to 310,000 gallons per day; a subsequent October 2025 city council presentation revised the figure to 350,000 gallons per day at full build-out; and a July 2026 report described, now air-cooled and closed-loop, as projected to use only 42,500 gallons per day, with the city imposing contractual penalties if usage exceeds 71,400 gallons on any single day~\cite{shawlocal-cardinal}. Under our framework, the water stewardship score would need to reflect not a single number but this trajectory, and would carry a self-reported confidence flag throughout, since none of these figures represent independently metered, publicly verified consumption; the city's daily-cap enforcement mechanism confirms a ceiling, not facilities' actual draw.

\textbf{Meta Data Center (DeKalb).} This facility offers the strongest available data-access precedent among the three. The city of DeKalb caps the facility at 200,000 gallons of water per day and meters its consumption directly, alongside every other residential and commercial water user in the city, with city staff reading the meter on a recurring basis~\cite{wglt-dekalb}. This places Meta DeKalb's water stewardship metric at a verified confidence level unattainable for the other two facilities in this sample, illustrating that recurring third-party metering, not merely a stated permit cap, is what separates a verifiable score from an estimated one. Although Meta reports annual facility-level water withdrawals, municipal meter readings and higher-frequency data remain unavailable. City metering verifies compliance with the cap, but independent assessment of consumption trends remains limited, which our local disclosure category captures.

raw municipal meter readings, temporal resolution, consumption versus withdrawal, and independently auditable telemetry.

\textbf{CyrusOne (Sangamon County).} This 280-acre, \$500 million, 600-megawatt facility was approved by a narrow 17-10 county board vote in April 2026 after months of contested hearings~\cite{wglt-sangamon}. Public records here are strongest on regulatory process (board votes, meeting minutes, vote tallies by member) and weakest on water and energy metrics: the developer has stated an intention to use a closed-loop cooling system to reduce water demand, but no permitted daily withdrawal cap, metering arrangement, or renewable energy commitment comparable to DeKalb's or Yorkville's has yet been made public at the time of writing, as the project still requires additional permits before construction~\cite{broadbandbreakfast-sangamon}. Under our framework, most categories for this facility would currently score as estimated or unavailable rather than verified or even self-reported, not because the facility's impact is necessarily worse than the other two, but because the public record has not yet reached the permitting stage where such figures are typically disclosed.

\textbf{Summary.} Table 3 sketches the confidence level available for the water stewardship category alone across all three sites, since it was the most consistently discussed metric in public records for each. Even for this single, comparatively well-covered category, no facility in our sample achieves a level of transparency sufficient for an independent party to construct a verified composite score without a formal, contracted audit relationship that grants direct access to metering data. This finding, that public records alone are insufficient to complete the framework we propose, is itself a central result of this illustrative application, and directly motivates the case for the recurring, contracted proposal audit mechanism. 

\begin{table}[h]
\centering
\vspace{-10pt}\caption{Proposed audit framework categories and weights} \vspace{-10pt}
\renewcommand{\arraystretch}{1.0}
\begin{tabular}{
  >{\raggedright\arraybackslash}p{2.0cm}
  >{\raggedright\arraybackslash}p{2.8cm}
  >{\raggedright\arraybackslash}p{2.5cm}
}
    \toprule
    \textbf{Facility} & \textbf{Public data available} & \textbf{Confidence level} \\
    \midrule
    Project Cardinal (Yorkville) & Multiple revised projections; contractual daily cap & Self-reported \\ \hline
    Meta (DeKalb) & Permitted cap; city-metered readings & Verified (cap only; totals undisclosed) \\\hline
    CyrusOne (Sangamon Co.) & Stated cooling approach; no cap or metering yet public & Estimated / unavailable \\
    \bottomrule
  \end{tabular}\vspace{-15pt}
\end{table}

\section{Verification, Independence, and Policy}
Existing data center sustainability metrics, PUE, WUE, and CUE chief among them, answer a facility-internal question: how efficiently does a data center convert resources into IT energy use. The disputes surveyed in this paper, Champaign County's moratorium~\cite{routefifty2026}, Yorkville's repeatedly revised water projections~\cite{weslo2025yorkville}, DeKalb's metered permit cap~\cite{howell2026aurora}, Sangamon County's contested approval~\cite{mackie2026sangamon}, and Illinois's stalled POWER Act~\cite{thorsberg2026illinois}, are organized around a different, place-specific question: what does this facility take from this particular community, and can that claim be verified. Our five-category framework answers that second question directly, weighted toward the water stewardship and local disclosure categories that existing instruments omit, and structured for recurring rather than one-time application.

The case studies in Section 4 expose why answering it is hard: this is a structural barrier, not an incidental gap. Outside DeKalb's metered arrangement~\cite{howell2026aurora}, none of our energy, water, or carbon/ renewables metrics are independently verifiable from public records alone. Regulatory compliance data is a public record floor available everywhere we looked, but everything above it, actual consumption trends, renewable procurement structure, and heat reuse, depends on voluntary operator disclosure or a formal audit relationship with contractual data-access rights. This is not a limitation specific to our framework; it reflects a broader condition in which a persistent lack of transparency leaves regulators and communities guessing about actual resource use. That public records alone cannot complete our own framework is as important a finding as the framework itself, and it is what motivates a recurring, contracted audit mechanism rather than a voluntary reporting standard: closing this gap requires either operator transparency at a scale not currently practiced, or the kind of standing, contractual obligation this paper's audit mechanism is designed to create.

This is also where the framework has the clearest practical effect: written into permitting conditions, not offered as a voluntary standard. Champaign County's moratorium will require a decision framework when it comes up for review~\cite{routefifty2026}; a structured audit requirement, rather than blanket approval or extension, offers a middle path that neither forecloses development nor leaves future applicants unaccountable to the concerns that motivated the moratorium. At the state level, the failed POWER Act would have mandated quarterly reporting on water use and renewable sourcing~\cite{thorsberg2026illinois}; our framework could serve as a candidate methodology for exactly such a mandate.

While our illustrative application focuses on Illinois, the pattern it exposes is not local. The same dynamics, unpermitted or self-reported water figures, permitting bodies negotiating without independently verified numbers, communities organizing around questions no existing metric answers, recur across the more than \$156 billion in projects delayed or canceled nationally in 2025 and the activist groups now organizing across two dozen states~\cite{datacenterwatch-q3q4-2025, datacenterwatch2026}. Illinois's moratorium wave and stalled POWER Act are not an anomaly; they are an early, unusually well-documented instance of a dispute now playing out with the same structure, and often the same missing data, in dozens of other jurisdictions. A framework validated at one site is of limited use to a county considering its own permitting conditions; a framework whose categories, weights, and maturity levels are explicit and published, as ours are, can be adopted or adapted by any jurisdiction facing the same fact pattern without requiring each community to rediscover the same data-access barriers independently.

\vspace{-15pt}
\section{Limitations}
Two limitations bound this contribution. First, the framework audits only on-site water and energy use; it does not yet capture indirect water consumption at the point of power generation~\cite{macknick2011}, a gap the POWER Act explicitly raised and one that can be substantial for facilities drawing from thermoelectric sources. Extending water stewardship to a facility's electricity mix is a natural next step. Second, the categories and weights proposed here still need validation against operator-provided metering obtained through a contracted audit relationship, rather than the public approximations this paper relies on. We offer this framework as a starting point for sustainable computing and for county and state policymakers to test against real permitting and disclosure requirements going forward.



\bibliographystyle{ACM-Reference-Format}
\bibliography{0.main}

\end{document}